\documentclass[conference]{IEEEtran}
\IEEEoverridecommandlockouts
\usepackage{cite}
\usepackage{amsmath,amssymb,amsfonts}
\usepackage{graphicx}
\usepackage{textcomp}
\usepackage{xcolor}
\usepackage{url}
\usepackage{bm}
\usepackage{multirow}
\usepackage{subfig}
\usepackage{booktabs}

\newcommand{\bs}[1]{{\mathbf{#1}}}
\usepackage[algo2e]{algorithm2e}
\usepackage{algorithm}
\begin{document}

\title{Efficient Batch Homomorphic Encryption for Vertically Federated XGBoost
}

\author{\IEEEauthorblockN{Wuxing Xu}
\IEEEauthorblockA{\textit{Beihang University}\\
Beijing 100191, China \\
xuwx@buaa.edu.cn}
\and
\IEEEauthorblockN{Hao Fan}
\IEEEauthorblockA{\textit{JD Technology Group}\\
Beijing 100101, China \\
fanhao26@jd.com}
\and
\IEEEauthorblockN{Kaixin Li}
\IEEEauthorblockA{\textit{Zhejiang University}\\
Hangzhou 310058, China \\
21921193@zju.edu.cn}
\and
\IEEEauthorblockN{Kai Yang}
\IEEEauthorblockA{\textit{JD Technology Group}\\
Beijing 100101, China \\
yangkai188@jd.com}
}

\maketitle

\begin{abstract}
More and more orgainizations and institutions make efforts on using external data to improve the performance of AI services. To address the data privacy and security concerns, federated learning has attracted increasing attention from both academia and industry to securely construct AI models across multiple isolated data providers. In this paper, we studied the efficiency problem of adapting widely used XGBoost model in real-world applications to vertical federated learning setting. State-of-the-art vertical federated XGBoost frameworks requires large number of encryption operations and ciphertext transmissions, which makes the model training much less efficient than training XGBoost models locally. To bridge this gap, we proposed a novel batch homomorphic encryption method to cut the cost of encryption-related computation and transmission in nearly half. This is achieved by encoding the first-order derivative and the second-order derivative into a single number for encryption, ciphertext transmission, and homomorphic addition operations. The sum of multiple first-order derivatives and second-order derivatives can be simultaneously decoded from the sum of encoded values. We are motivated by the batch idea in the work of BatchCrypt for horizontal federated learning, and design a novel batch method to address the limitations of allowing quite few number of negative numbers. The encode procedure of the proposed batch method consists of four steps, including shifting, truncating, quantizing and batching, while the decoding procedure consists of de-quantization and shifting back. The advantages of our method are demonstrated through theoretical analysis and extensive numerical experiments.
\end{abstract}


\section{Introduction}
The potential of data are increasingly explored in a variety of areas, for which more and more organizations and institutions are seeking for external data providers to construct high-quality AI models such as credit card fraud detection \cite{yang2019ffd}, page recommendation \cite{zhao2020privacy}, and health care \cite{rieke2020future}. However, it is usually undesireble and even not legitimate to directly share original data with other institutions, for maintaining the ownership of high-value data and protecting the privacy of their users' data. Many regulations have been enacted to protect the data privacy and security, including General Data Protection Regulation (GDPR) by European Union \cite{GDPR}, and the recently passed China's Personal Information Protection Law\footnote{\url{http://www.mod.gov.cn/regulatory/2021-08/20/content_4892505.htm}}. As a representative, the data of loan facilitation institution are not allowed to be shared with financing institution directly for consumer loan in China.
\begin{figure*}[h]
    \centering
    \includegraphics[width=0.7\textwidth]{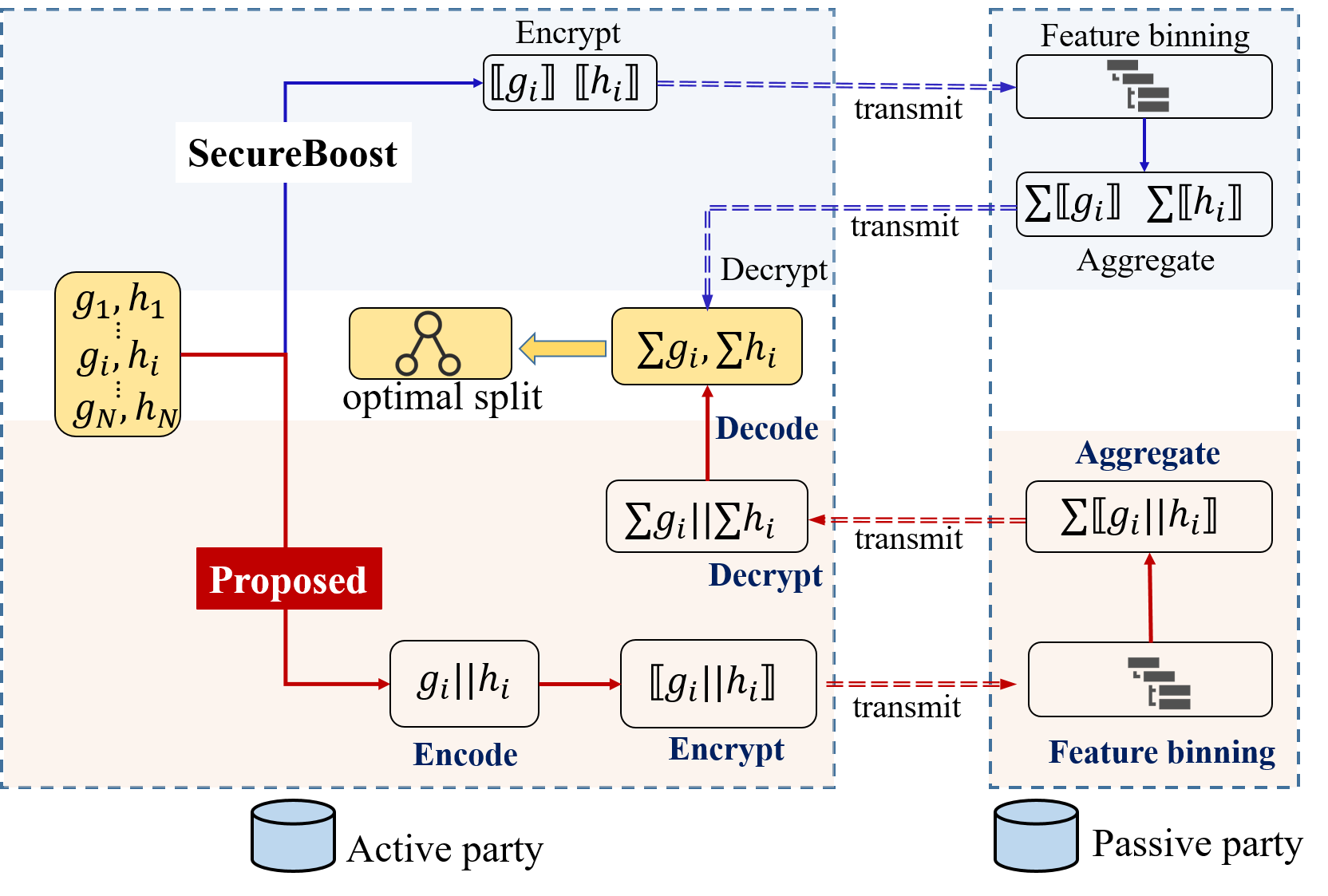}
    \caption{The framework of proposed efficient batch homomorphic encryption method for vertical federated XGBoost compared with state-of-the-art algorithm SecureBoost.}
    \label{fig:framework}
\end{figure*}

To exploit the full potential of cross-silo data sources, the emerging field termed as \textit{federated learning} \cite{mcmahan2017communication,yang2019federated} has garnered numerous attention for studying privacy-preserving machine learning methods while leaving original data locally at each provider. According to the distributed data structure, i.e., horizontally partitioned data or vertically partitioned data, federated learning can be categorized into \textit{horizontal federated learning}, and \textit{vertical federated learning}. Horizontal federated learning area studies the applications where each node has access to a subset of data instances with common features \cite{mcmahan2017communication,li2020privacy}, such as online shopping behaviors for users from different cities. Vertical federated learning studies the cases when each institution owns a subset of features for common users \cite{cheng2019secureboost,gu2020federated,yang2019quasi}, such as online shopping behaviors from an e-commercial company and credit card usage information from a bank. 

Vertical partitioned data structure is common and of particular interest in real-world applications of cross-institution cooperations. An institute usually has already owned data features of specific domains related to its main business, and labels as the target of the AI model to be constructed (often called model used party or active party). However, the performance of self-owned data is not satisfactory and can be further improved by external data from other domains (often called data provider or passive party).
Distinct from horizontal partitioned data, in vertical federated learning, 
participating parties accomplish model training interactively, by performing computations and exchanging (often encrypted) intermediate values without revealing original data \cite{hardy2017private,yang2019quasi,cheng2019secureboost,feng2019securegbm,gu2020federated,liu2020federated}. For example, in vertically federated linear models, a gradient is divided into local terms that can be computed at each party, and cross term that can be computed by sending the encrypted intermediate value from one participating party to another \cite{hardy2017private,yang2019quasi}. For vertically federated XGBoost models, it is critical to compute the gain and weight for each possible split point of data provider without revealing labels owned by model used party. The work of SecureBoost \cite{cheng2019secureboost} found that both gain and weight are functions of aggregated gradients $g$ and $h$, and thus proposed to interactively compute aggregated gradients with additively homomorphic encryption (HE) to build vertical federated XGBoost models.

In this paper, we focus on vertical federated XGBoost model due to the powerful generalization capability, training efficiency, interpretability and the widely use in real-world applications \cite{cheng2019secureboost,feng2019securegbm}. Unfortunately, large number of encryption operations and ciphertext transmissions make the training of vertical federated XGBoost models much less efficient than local versions. For example, the well-known paritially HE system Paillier \cite{paillier1999cryptosystems} takes about 100s to encrypt 10000 intergers (11 KB in total) with a 1024bit key (average over $5$ realizations conducted on a laptop with Intel i7-10510U CPU), and the length of the ciphertext is about 30 times larger than that of the plaintext for single encryption operation. To reduce the large amount of overhead resulted by HE in vertical federated XGBoost, in this paper, we propose to encoding the first-order derivative $g_i$ and second-order derivative $h_i$ to a single value for encryption and transmission, which saves almost half the cost, as illustrated in Figure \ref{fig:framework}. 

This batching method is applicable based on the key observation that the first and second derivatives are operated in the same way after being encrypted. Such batching idea is motivated by the work of BatchCrypt \cite{zhang2020batchcrypt}, which compresses the high-dimensional gradient of deep neural network model into much lower dimensions to reduce the HE cost in horizontal federated learning. However, directly exploiting the principles of BatchCrypt in vertical federated XGBoost will lead to overflow errors frequently. We find that the overflow errors occurred due to adding negative numbers represented in two's complement. To address this issue, we propose a novel batch method by shifting all to non-negative numbers, truncating, quantizing them to unsigned binary numbers, and putting them together as a batched number for further encryption, transmission, and computations. 

\textbf{Our contributions.} We summarize the contributions of this paper in the following
\begin{itemize}
	\item We studied the vertical federated learning problem of XGBoost model assuming all parties have distinct features for a common set of users, where one active party has access to a subset of features and labels, and at least one passive party has access to a different subset of features. We identified the efficiency problem resulted from frequent homomorphic encryption computations and transmissions, which is the key limiting factor for real-world applications.
	\item We proposed to leverage the batching idea to reduce the number of homomorphic encryption computations and transmissions by half via encoding a first-order derivative and a second-order derivative into a single number. 
	\item We provided an analysis of the selection of hyperparameters and their impacts on the precision and possible overflow, which makes our batch method quite handy in applications.
	\item We conducted extensive numerical experiments to evaluate the performance of our method and it without batching. Numerical results demonstrated that the proposed method can significantly improve the efficiency of vertical federated XGBoost model training by up to $44\%$ and reduce the total runtime from 85 minutes to 48 minutes.
\end{itemize}
\textbf{Organization.} The rest of this paper is arranged as follows. In Section II, We review the XGBoost algorithm and state-of-the-art vertical federated XGBoost frameworks, followed by stating the efficiency problem resulted by frequent homomorphic encryption operations. In Section III, we present our efficient batch method aided vertical federated XGBoost framework. In Section IV, we evaluate the proposed algorithm and benchmarks using extensive numerical experiments. Finally, Section V concludes this paper.

\section{Preliminary Studies and Problem Statement}
In this section, we provide a brief introduction on the XGBoost model, followed by state-of-the-art vertical federated XGBoost frameworks. The critical efficiency concerns resulted by the extremely high overhead for encryption operations and ciphertext transmissions are then analyzed.

\subsection{XGBoost}
Gradient boosting is a powerful and widely applied machine learning framework by ensembling a number of base learners. Under gradient boosting framework, XGBoost \cite{cheng2019secureboost} stands out as a highly efficient and flexible implementation.

Specifically, given a dataset $D = \{(\bs{x}_i; y_i); \bs{x}_i \in \mathbb{R}^{d},y_i\in\mathbb{R},0<i\leq N \} $ contains $N$ data instances and $d$ features, the model's prediction output for a certain data $\bs{x}_i$ in the data set is composed of the results of $K$ regression trees as follows: 
\begin{equation}
\hat{y}_i=\sum^K_{t=1}f_t(\bs{x}_i).
\end{equation}

XGBoost adopts an additive training process. At the $t$-th iteration, a tree $f_t$ is generated to minimize a second-order approximation loss function to improve model training efficiency, which is given by
\begin{equation}
\mathcal{L}^{(t)} \simeq \sum^n_{i=1}[l(y_i,\hat{y}^{(t-1)})+g_if_t(\bs{x}_i)+\frac{1}{2}h_if_t^2(\bs{x}_i)]+\Omega(f_t).
\end{equation}
$\Omega(f_t) = \gamma T +\frac{1}{2}\lambda \sum_{j=1}^{T}w_j^2$ is the regularization term, where $T$ is the number of leaf nodes and $w_j$ is the weight score of $j$-th leaf node. $g_i= \partial _{\hat{y}^{t-1}}l(y_i, \hat{y}^{(t-1)}) $ and $h_i= \partial^{2} _{\hat{y}^{t-1}}l(y_i, \hat{y}^{(t-1)}) $ are the first and second
derivative of the loss function at $\hat{y}^{(t-1)}$, respectively.

The decision whether a node will be split is made according to the gain given by
\begin{equation}\label{eq:gain}
{\sf{Gain}}=  \frac{1}{2}[
\frac{(\sum_{i\in I_L}g_i)^2}{\sum_{i\in I_L}h_i+\lambda}+
\frac{(\sum_{i\in I_R}g_i)^2}{\sum_{i\in I_R}h_i+\lambda}-
\frac{(\sum_{i\in I}g_i)^2}{\sum_{i\in I}h_i+\lambda}]-\gamma.
\end{equation}
The regularization parameter $\gamma>0$ controls the model complexity, such that only nodes with gain greater than $\gamma$ will be split into left and right child nodes. The optimal split point is selected by maximizing gain among all possible split points for each feature. When the maximum depth is reached, the optimal weight $w_j$ of the $j$-th leaf node will be computed as
\begin{equation}\label{eq:leafweight}
w_j = -\frac{\sum_{i\in I}g_i}{\sum_{i\in I}h_i+\lambda},
\end{equation}
where $I$ represents the instance space of the leaf node.

\subsection{Vertical Federated XGBoost}\label{subsec:verticalFedXGB}
In vertical federated learning, we study the case when a number of parties hold different features. In real-world applications, it is of particular interest that only one party has access to labels, which is called the \textit{Active Party}. The Active Party wants to use the features held by other parties, called \textit{Passive Party} to improve the performance of its machine learning model. 
\begin{itemize}
\item {\itshape{\bfseries Active Party (AP).} The model used party, who holds both a data matrix and the class label, wants to use the passive party's data to improve the performance of AI models.}

\item {\itshape{\bfseries Passive Party (PP).} The data provider party, who holds a data matrix without labels, will use its data to improve the performance of AI models required by Active Party. }
\end{itemize} 

Mathematically, we can represent the data held by the $p$-th ($1\leq p \leq P$) passive party as $D^{p}=\{\bs{x}_i^{p}\in\mathbb{R}^{d_p},1\leq i\leq n\}$ and the data held by the active party as $D^{0}=\{(\bs{x}_i^{0}\in\mathbb{R}^{d_0},y_i),1\leq i\leq n\}$. The distributed features held by each party $\{\bs{x}_i^{p}\}_{p=0}^{P}$ can thus be looked as the vertical split on one complete data instance $\bs{x}_i$. 

As one of the most well-known vertical federated XGBoost frameworks, the work of SecureBoost \cite{cheng2019secureboost} proposed to exchange derivatives $g_i$'s and $h_i$'s in ciphertext to securely compute the gain for node split with partially homomorphic encryption (HE), while leaving original data local at each party. The work of SecureGBM \cite{feng2019securegbm} further adapted LightGBM, a fast implementation of XGBoost, to vertical federated learning case. In vertical federated XGBoost, the decision whether each node is split or not is made according to the split gain. Since the label $y_i$ and model prediction value $\hat{y}_i$ are only available to the Active Party, gradients $g_i$ and $h_i$ can be directly computed by the Active Party locally. Therefore, the gain can also be computed locally for each possible split point of the Active Party's features. The key to vertical federated XGBoost is to compute the gain for each possible split point of the Passive Party's feature, and the leaf weight. Observing that both of gain and leaf weight are functions of aggregated first-order derivative $g_i$ and second-order derivative $h_i$, existing vertical federated XGBoost frameworks proposed to send all encrypted gradients $g_i$ and $h_i$ to Passive Parties and they compute aggregated gradients in ciphertext. By transmitting the encrypted aggregation gradients back, the Active Party can decrypt them and compute gain and leaf weights in plaintext.

Although partially HE is one of the most efficient methods to accomplish model training without revealing data privacy, the high computation overhead and transmission overhead are still main bottlenecks to make federated XGBoost as efficient as the local version. The design target of this paper is thus to improve the computation and communication efficiency of vertical federated XGBoost methods.

\subsection{Problems Analysis}
\textbf{Why is partially HE still slow in XGB?}
The overhead resulted from cryptographic system in vertical federated XGBoost consists of two parts, i.e., computation and transmission. The Active Party should encrypt $g_i$ and $h_i$ and transmit them to Passive Parties, and Passive Parties shall compute the sum for every possible split and transmit them back to the Active Party. Finally, the Active Party should obtain the plaintexts via decryption. 
In such one split process, the Active Party needs to encrypt $2n$ times and transfer $2n$ ciphertexts with $n$ samples in data set, which is the chief source of overhead. For example , the well-known partially HE system Paillier takes about 100s to encrypt 10000 intergers (11 KB in total) with a 1024-bit key (average over $5$ realizations conducted on a laptop with Intel i7-10510U CPU), and the length of the ciphertext is about 30 times larger than that of the plaintext for single encryption operation.
The high costs of encryption and ciphertext transmissions make vertical federated XGBoost methods much slower than local model training. 

In this paper, we aim to improve the computation and communication efficiency to further mitigate the inefficiency caused by HE in vertical federated XGBoost frameworks, which is achieved by reducing the amount of ciphertext calculation and transmission.
It is based on the key observation that $g_i$ and $h_i$ are operated in the same way in the splitting process, including encryption, transmission, aggregation, and decryption. The work of BatchCrypt \cite{zhang2020batchcrypt} motivates us to encode multiple scalars as a single scalar to reduce the number of encryption operations and transmissions.
In \cite{zhang2020batchcrypt}, the authors proposed BatchCrypt to compress high-dimensional gradients of deep learning model at each data provider, whose results are encrypted and transmitted to the centralized server to calculate the aggregation securely for horizontal federated learning. Unfortunately, it is inapplicable to directly adopt the BatchCrypt framework to compress $g$ and $h$ as one plaintext to improve the efficiency of vertical federated XGBoost.

\begin{figure}[h]
    \centering
    \includegraphics[width=\columnwidth]{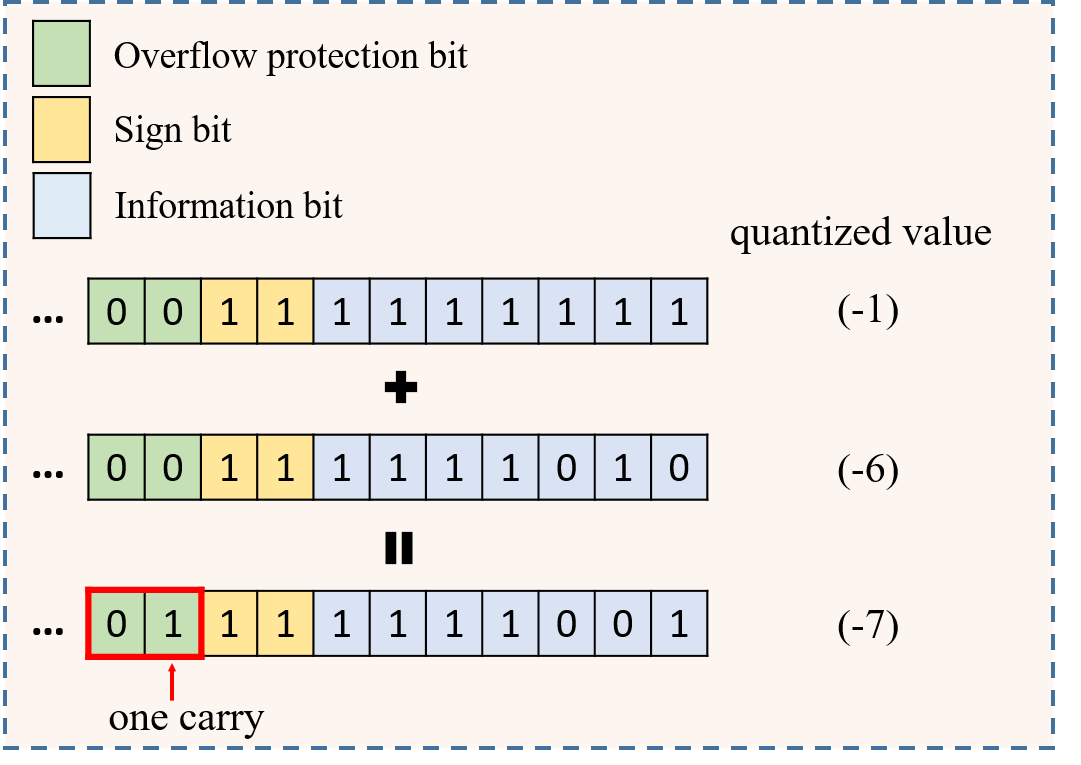}
    \caption{An example of negative overflow for BatchCrypt method.}
    \label{fig:example_batchcrypt}
\end{figure}

\textbf{Why does not BatchCrypt work for vertical federated XGBoost?} The main challenge is that the overflow error comes frequently, if not always. In BatchCrypt, a batch of scalars are truncated to $[-\alpha,\alpha]$, quantized into signed integers, and then encoded to a long integer and encrypted in one go. We find that overflow errors are caused by adding multiple negative values. Since the sign bit of a negative number in two’s complement representation is always 1, adding two negative numbers leads to arithmetic carry. For example, $-1$ plus $-6$ (quantized values) will turn overflow bits from `00' to `01' although the representation range of $6$ information bits and sign bits is [-127,127], as illustrated in Fig. \ref{fig:example_batchcrypt}. 
That is, the overflow may still occur even though the sum is still within the designed representation range. The overflow protection bits will become `11' as long as there are more than $4$ negative values in the summation with $2$ overflow protection bits, which will raise an overflow error.

This problem is much less common and thus rarely observed in the cross-institute federation learning, because there are many values to be batched but much fewer to be added (9 in the simulations of BatchCrypt). In vertical federated XGBoost, however, there are thousands of $g_i$'s and $h_i$'s to be aggregated for each possible split point. Therefore, in this work, we shall devise a novel batch method to improve HE computation and ciphertext communication efficiency of vertical federated XGBoost while avoiding the overflow of adding negative numbers.

\section{Proposed Efficient Batch Homomorphic Encryption Method}
In this section, we first introduce the principles of our batch method to avoid the overflow of adding negative numbers, followed by presenting the overall batch homomorphic encryption aided vertical federated XGBoost framework. 

\subsection{Proposed Batch Method}
The proposed batch method consists of three parts, i.e., encoding, aggregation, and decoding. The basic idea of batch method to avoid negative overflow is to shift all values to non-negative numbers and map multiple numbers as a single number. 

\subsubsection{Encoding}
Given $n$ vectors $\bs{m}_1,\cdots,\bs{m}_n\in\mathbb{R}^{d}$, we propose to encoding each vector $\bs{m}_i$ to a scalar $z_i\in\mathbb{R}$ following four steps for computing the (weighted) sum $\bs{m}_{\sf{sum}}=\sum_{i=1}^{n}\bs{m}_i$ securely, including shifting, truncating, quantizing, and batching.
\begin{itemize}
    \item \textbf{Shifting} $\bs{m}_{i}$ to non-negative number $\bs{u}_{i}$. 
    Each entry of $\bs{m}_i$, i.e., $m_{ij}$, need to be shifted to non-negative numbers $u_{ij}$ in $\bs{u}_i$ by subtracting a sufficient small number $s_{j}$ as follow
    \begin{equation}
    u_{ij} = x_{ij}-s_{j}.
    \end{equation}
    The shift $s_{j}\leq min (\{m_{ij}\}_{i=1}^{n},0)$ is no greater than the minimum value of each vector's j-th entry. 
    \item \textbf{Truncating} $\bs{u}_{i}$ to $[0,\alpha]$. 
    To prevent overflow caused by large numbers, we set a hyperparameter $\alpha$ and truncate the $j$-th entry of shifted vector $\bs{u}_i$ (denoted by $u_{ij}$) into range $[0,\alpha]$ for any $j=1,\dots,d$, which can be represented as 
    \begin{equation}
    u_{ij}=min(u_{ij},\alpha);~j=1,\dots,d.
    \end{equation}
    \item \textbf{Quantizing} $u_{ij}$ to $r$-bit number $q_{ij}$. Since all $u_{ij}$'s are non-negative after shifting, we can quantize them into unsigned numbers. Given parameter $\alpha_{\sf{max}}\geq \alpha$ for any $j=1,\dots,d$ and number of information bits $r$, we uniformly map $[0,\alpha_{\sf{max}}]$ to $[0, 2^r-1]$. Then each $u_{ij}$ will be mapped as an $r$-bit positive integer, which is given by
    \begin{equation}\label{eq:quantization}
    q_{ij}=\lfloor 2^r \cdot\frac{u_{ij}}{\alpha_{\sf{max}}}+\frac{1}{2}\rfloor.
    \end{equation}
    \item \textbf{Batching} $q_{ij}$ to an $(r+2)d$-bit number $z_i$. We put a batch of quantized values $q_{ij}$ for $j=1,\dots,d$ together as a scalar $z_i$ by inserting 2 overflow protection bits (can be any value no less than 1) between them. It can be represented as
    \begin{equation}
    z_i=\underbrace{\overbrace{[01101...]}^{r\mbox{-}bits}}_{q_{i1}}
    \underbrace{\overbrace{[00]}^{2\mbox{-}bits}}_{padding} \underbrace{\overbrace{[10010...]}^{r\mbox{-}bits}}_{q_{i2}}\cdots
    \underbrace{\overbrace{[00]}^{2\mbox{-}bits}}_{padding} \underbrace{\overbrace{[10001...]}^{r\mbox{-}bits}}_{q_{id}}.
    \end{equation}
\end{itemize}

\subsubsection{Aggregation}
We can further perform addition on the batching number, i.e., $z_1+z_2$. Partially homomorphic encryption such as Pailler enables us to perform addition operations on ciphertext, which can be represented as $Enc(z_1)\oplus Enc(z_2) = Enc(z_1+z_2)$. $\oplus$ and $\odot$ denote the homomorphic addition and dot product operation, respectively. Therefore, the aggregated result $z_{\sf{sum}}=\sum_{i=1}^{n}z_i$ can be computed securely by sending the encrypted value of each $z_i$ to another party and obtaining $Enc(z_{\sf{sum}})=\sum_{i=1}^{n}Enc(z_i)$. 

\subsubsection{Decoding} The decoding procedure of the aggregated vector $\bs{x}_{\sf{sum}}=\sum_{i=1}^{n}\bs{x}_i$ from the aggregated value $z_{\sf{sum}}$ is given by the following two steps:
\begin{itemize}
\item \textbf{De-quantization} to the shifted sum. Firstly, we can obtain a batch of $q_{{\sf{sum}},j}=\sum_{i=1}^{n}q_{ij}$ following
\begin{equation}
q_{{\sf{sum}},j}=z_{{\sf{sum}},j}\cdot\frac{\alpha_{\sf{max}}}{2^r}, \forall j=1,\dots,d,
\end{equation}
where $z_{{\sf{sum}},j}$ is extracted from
\begin{equation}
z_{\sf{sum}}=\underbrace{\overbrace{[0001101...]}^{(r+2)\mbox{-}bits}}_{z_{\sf{sum},1}}\underbrace{\overbrace{[0010010...]}^{(r+2)\mbox{-}bits}}_{z_{\sf{sum},2}}\cdots\underbrace{\overbrace{[0010001...]}^{(r+2)\mbox{-}bits}}_{z_{\sf{sum},d}}.
\end{equation}
\item \textbf{Shifting back.} We can then obtain the estimated value of each entry of the aggregated vector $\hat{\bs{m}}_{\sf{sum}}$ by shifting back, i.e.,
\begin{equation}
\hat{m}_{\sf{sum},j}=q_{{\sf{sum}},j}+n\cdot s_{j}.
\end{equation}
\end{itemize}
Note that when the overflow protection bits are `11', the decoded result is not reliable and an overflow error should be raised. We will provide a comprehensive analysis for the selection of hyperparameters in Section \ref{subsec:analysis}.

\begin{figure}[h]
    \centering
    \includegraphics[width=\columnwidth]{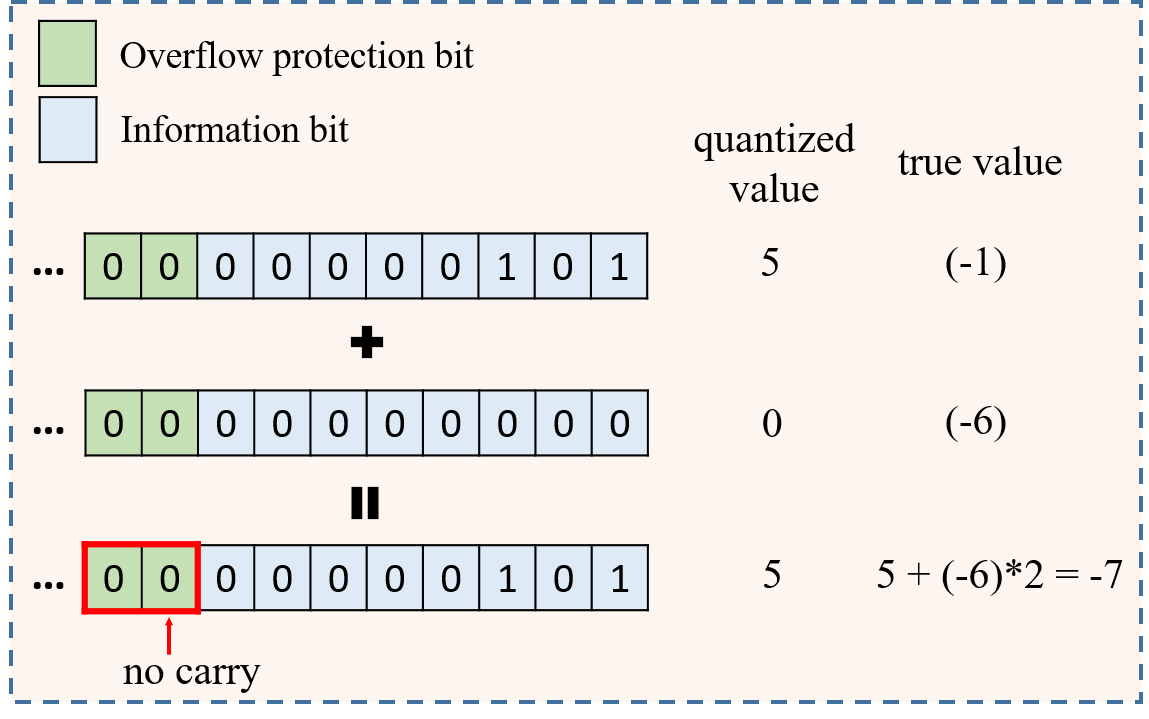}
    \caption{An example of avoiding negative overflow for proposed batch method.}
    \label{fig:example_proposed}
\end{figure}

In order to better illustrate our scheme, we still take -1 plus -6 as examples. We first shifted these two numbers into non-negative numbers by subtracting $s_j=-6$. By setting the truncation parameter $20$ and $\alpha_{\sf{max}}=127$, the shifted numbers $5$ and $0$ can be quantized as 8-bit binary number `00000101' and `00000000' losslessly, as shown in Figure \ref{fig:example_proposed}. The sum `00000101' can be decoded as $-7$ via de-quantization and shifting back by subtracting $2\cdot (-6)$, where the overflow protection bits remain `00'.

\subsection{Efficient Batch HE for Vertical Federated XGBoost}
After clarifying our novel batching method for avoiding negative overflow, here we present the overall procedure of our efficient batch HE method for vertical federated XGBoost. 

As pointed in Section \ref{subsec:verticalFedXGB}, the key to federated design of XGBoost model is the calculation of split gain and leaf weight. Firstly, the Active Party computes first-order and second-order derivatives $g_i, h_i$ of each data instance based on label $y_i$ and prediction value $\hat{y}_i^{(t-1)}$ in the $t$-th iteration. The Active Party then encodes all $g_i, h_i$'s into batched gradients $z_i$ using the proposed batching method, encrypts them with partially HE such as Paillier, and sends the encrypted values $Enc(z_i)$ to each Passive Party. In the following, the $t$-th tree model is built by repeatedly node split until the split gain given by equation (\ref{eq:gain}) of all leaf nodes is below $0$. The weight of all leaf nodes are then computed. 

For feature $k$ held by a the $p$-th party, the data $\{x_{k,i}^{p}\}_{i=1}^{n}$ can be divided into $L$ buckets denoted by $\mathcal{B}_1,\dots,\mathcal{B}_L$ in ascending order by $L-1$ split points $b_1\leq\cdots\leq b_{L-1}$, i.e., $x_{k,i}\leq b_{\ell}\leq x_{k,j} $ for all $i\in\mathcal{B}_{\ell},j\in\mathcal{B}_{\ell+1}$. For each split point $b_i$ of a node with instance space $\mathcal{I}$, the instance space of left child node is given by $\mathcal{I}_L=\left(\cup_{\ell=1}^{i}\mathcal{B}_{\ell}\right)\cap \mathcal{I}$, and the instance space of right child node is given by $\mathcal{I}_R=\left(\cup_{\ell=i+1}^{L}\mathcal{B}_{\ell}\right)\cap \mathcal{I}$.
\begin{itemize}
	\item Gain of features held by Active Party. The Active Party directly can compute the aggregated first-order derivatives and second-order derivatives in $\mathcal{I}_L$ and $\mathcal{I}_R$, which is given by $g_{\sf{sum},L}=\sum_{i\in\mathcal{I}_{L}}g_i$, $g_{\sf{sum},R}=\sum_{i\in\mathcal{I}_{R}}g_i$, $h_{\sf{sum},L}=\sum_{i\in\mathcal{I}_{L}}h_i$, and $h_{\sf{sum},R}=\sum_{i\in\mathcal{I}_{R}}h_i$, respectively. We can thus compute the gain for each split point and find the best Active Party's split point with maximum gain ${\sf{Gain}}^{\sf{AP}}$.
	\item Gain of features held by Passive Parties. Each Passive Party should compute the aggregated batched gradient in ciphertext within each bucket given the instance space $\mathcal{I}$ of each node, which is given by $Enc(z_{\sf{sum},\ell})=\sum_{i\in\mathcal{B}_{\ell}\cap \mathcal{I}}Enc(z_i)$. The aggregated batched gradients in ciphertext are transmitted to the Active Party, which are decrypted and decoded as aggregated first-order and second-order derivatives $g_{\sf{sum},\ell}$ and $h_{\sf{sum},\ell}$, respectively. The Active Party can then compute the gain for each split point and find the best Passive Party's split point with maximum gain ${\sf{Gain}}^{\sf{PP}}$.
\end{itemize}
Therefore, the Active Party obtains the best split point and maximum gain ${\sf{Gain}}=\max({\sf{Gain}}^{\sf{AP}},{\sf{Gain}}^{\sf{PP}})$. If the maximum gain is above zero, the node will be split to left and right child node with instance space $\mathcal{I}_L$ and $\mathcal{I}_R$, respectively. Otherwise, the node will not be split any more and the corresponding leaf weight is given by equation (\ref{eq:leafweight}). 

We summarize the overall training procedure of our batch HE aided vertical federated XGBoost method in Algorithm \ref{algorithm:proposed}.

\SetNlSty{textbf}{}{:}
\IncMargin{1em}
\begin{algorithm}[htb]
\SetKwInOut{Input}{Input}
\SetKwInOut{Output}{Output}
\Input{Aligned vertically split $N$ data instances at each party $p=0,\dots,P$ via private set intersection \cite{liang2004privacy}}
\Output{$T$ constructed boosting tree model $f_1,\dots,f_T$}\vspace{0.5em}

\textbf{Key generation:} The AP generates a Paillier key pair, and broadcasts the public key to each PP\\
\textbf{Feature binning:} Each party $j$ divide its own feature $k$ as $L$ buckets $\mathcal{B}_1,\dots,\mathcal{B}_L$ and obtain the splits points as candidates for each node splitting.\\
\textbf{Building boosting trees} Construct $T$ trees sequentially\\
\For{$t=1,\cdots,T$}{
\textbf{Prepare gradients}\\
AP: Computing gradients $g_i$ and $h_i$ based on $y_i$ and $\hat{y}_i^{(t)}$. \\
AP: Encoding each pair of $g_i$ and $h_i$ into $z_i$ and sending the encrypted values to all PPs.\\
\textbf{Iteratively nodes split}\\
Inititialize the tree whose node set $\mathcal{N}=\{\text{Node}_0\}$ has only a root node with instance space as $\mathcal{I}=1,\dots,N$ \\
\While{$\mathcal{N}\ne \emptyset$}{
Choose any $j\in\mathcal{N}$\\
\textbf{Computing gain of AP's features}\\
AP: directly computes the split gain following equation (\ref{eq:gain}) as ${\sf{Gain}}^{\sf{AP}}$

\textbf{Computing gain of PPs' features}\\
PP: Each PP computes the encrypted aggregated batched gradients within each bucket $\mathcal{B}_{\ell}\cap\mathcal{I}$ and sends back to the AP \\
AP: decrypts and decodes each aggregated batched gradients, and computes the maximum split gain for PPs as ${\sf{Gain}}^{\sf{PP}}$.

\uIf{${\sf{Gain}}=\max({\sf{Gain}}^{\sf{AP}},{\sf{Gain}}^{\sf{PP}})>0$}{split node $j$ and add child nodes to $\mathcal{N}$}
\Else
{
 compute the leaf weights following equation (\ref{eq:leafweight}) and remove node $j$ from $\mathcal{N}$.
} 
}
$f_t$ has been constructed. Computing the prediction value of each data instance $\hat{y}_i^{(t)}=\sum_{i=1}^{t}f_i(\bs{x}_i)$.
}
Obtain the constructed federated XGBoost model $f_{1},\dots,f_{T}$.\\
\caption{Proposed Batch Method Aided Vertical Federated XGBoost.}
\label{algorithm:proposed}
\end{algorithm}

\subsection{Analysis and discussion}\label{subsec:analysis}
The proposed batch method significantly reduces the computation and communication cost of HE while introducing loss of precision due to truncation and quantization. In our proposed batch method, there are a list of hyperparameters, including the shifting parameter $s_j$'s, truncating parameter $\alpha$, and quantization parameter $\alpha_{\sf{max}}$ and $r$. It is important is to analyze how to select hyperparameters to avoid overflow and keep high precision. In this subsection, we will answer the questions with detailed analysis on our batch method. 

\textbf{How to avoid overflow error?} Here we claim that the overflow error occurs when the overflow protection bits becomes `11', i.e.,
\begin{equation}
 	\sum_{i=1}^{n}q_{ij}\geq \text{binary2decimal}(11\underbrace{000\cdots}_{r-bits})=2^r+2^{(r+1)}.
 \end{equation} 
Since the number before quantization is truncated to $[0,\alpha]$, we can obtain
\begin{equation}
	\sum_{i=1}^{n}q_{ij}\leq \sum_{i=1}^{n} \Big(2^r \cdot\frac{\alpha}{\alpha_{\sf{max}}}+\frac{1}{2}\Big)=n\cdot\Big(2^r\frac{\alpha}{\alpha_{\sf{max}}}+\frac{1}{2}\Big).
\end{equation}
Therefore, a sufficient condition of no overflow error for the proposed method is given by
\begin{equation}
	n\cdot\Big(2^r\frac{\alpha}{\alpha_{\sf{max}}}+\frac{1}{2}\Big) < 2^r+2^{(r+1)}.
\end{equation}
For simplicity in choosing parameters, we give a looser condition (not exactly sufficient) such that the overflow protection bits keeping `00' as follow
\begin{equation}
	n\alpha<\alpha_{\sf{max}}.
\end{equation}

\textbf{What is the precision?} The proposed batch method is lossy since we adopt lossy operations including truncation and quantization during batch encoding. 
The loss measured by accumulated error is given by
\begin{align}
	\text{loss}_{\sf{err}}&=\sum_{i=1}^{n}\left(q_{ij}\frac{\alpha_{\sf{max}}}{2^r}+s_j-m_{ij}\right) \nonumber\\
	&=\sum_{i=1}^{n}\underbrace{(q_{ij}\frac{\alpha_{\sf{max}}}{2^r}-\min(u_{ij},\alpha))}_{\text{quantization error}}-\underbrace{\left(u_{ij}-\min(u_{ij},\alpha)\right)}_{\text{truncation error}}. 
	\end{align}
Since error of rounding-to-nearest falls within the range of $[-1/2,1/2]$ and the truncation error is given by $\max(0,u_{ij}-\alpha)$, the loss measured by accumulated abosolute error $\text{loss}_{\sf{abs\_err}}$ is given by
\begin{align}
	\!\!\!\text{loss}_{\sf{err}}\leq \text{loss}_{\sf{abs\_err}}&=\sum\left|\text{quant. error}\right|+\left|\text{trunc. error}\right|\nonumber\\
	&\leq \frac{n}{2}\cdot\frac{\alpha_{\sf{max}}}{2^r}+\sum_{i=1}^{n}\max(0,u_{ij}-\alpha),
	\end{align}
where $\alpha_{\sf{max}}/2^r$ is termed as the resolution of quantization.

Note that the bound of quantization loss is calculated under round-to-nearest strategy as indicated in equation (\ref{eq:quantization}). We can also adopt stochastic rounding strategy to stochastically preserve diminishing information \cite{wen2017terngrad}, which is beyond the scope of this paper. Through numerical experiments, we find that the loss of precision is already acceptable and the performances are comparable with the non-batching method. 

There are two tradeoffs between precision and avoiding overflow. Firstly, increasing $\alpha_{\sf{max}}$ will allow more number of values to be aggregated without overflow (i.e., $n$), while the precision will decrease since the quantization error increases. Secondly, decreasing $\alpha_{\sf{max}}$ will also allow more number of values to be aggregated without overflow, while the precision will still decrease since the truncation error increases. 

\textbf{Analysis on computation and communication cost.} Since we encode first-order and second-oerder derivatives into a single number, the proposed batch HE aided vertically federated learning approach only requires half the number of encryption operations, homomorphic addition operations, and network traffic of ciphertext as it without batching \cite{cheng2019secureboost,feng2019securegbm}. Moreover, the batching method is parallel and compatible with other speeding-up techniques, such as Gradient-based One-Side Sampling (GOSS) \cite{feng2019securegbm,ke2017lightgbm}. We increase the computation cost by additional encoding operations on plaintext, and the communication cost by additional number of aggregation values $n$, which are much cheaper than encryption-related computation and transmission. 


\section{EXPERIMENTS AND RESULTS}
In this section, we conduct extensive numerical experiments to evaluate the performance of state-of-art methods.

\subsection{Experimental Setup}
The algorithms used in experiments are introduced below:
\begin{itemize}
	\item \textbf{``SecureBoost''} \cite{cheng2019secureboost}: We choose the well-known vertical federated XGBoost framework, SecureBoost, as the benchmark of non-batching method, which is implemented in FATE \cite{webankfate}. We choose the FATE 1.5 version in our experiments. 
	\item \textbf{``Proposed''} : We implement our batch method by revising the code of SecureBoost in FATE 1.5 to guarantee the fairness of comparisons.
\end{itemize}
We consider the vertical federated learning system with one Active Party and one Passive Party, and compare the performances of different approaches on the following datasets:
\begin{itemize}
\item \textit{Give Me Some Credit \cite{dataset:givecredit}} (termed as \textbf{``Giveme''}): It is an open-access credit dataset consisting of 150000 data instances with 10 features in total.
\item \textit{Default of Credit Card Clients \cite{dataset:defaultcredit}} (termed as \textbf{``Default''}):
It is another open-access credit scoring dataset involving the data of 30000 instances and each instance has 25 attributes, which is correlated to the classification task of predicting whether a user would repay on time.
\end{itemize}
We simulate a vertical federated learning setting by splitting the features for each dataset into two parts and deployed at one Active Party and one Passive Party, respectively. The detailed information is listed in Table \ref{tab:datatable}.
\begin{table}[h]
\centering
\caption{Datasets used in experiments.}
\begin{tabular}{ccccc}
\toprule
  \multirow{2}{*}{Datasets} & \multirow{2}{*}{Train Samples} & \multirow{2}{*}{Test Samples} & \multicolumn{2}{c}{Features} \\
  & & & Active Party & Passive Party\\
\midrule
Default & 18300 &  5699 & 5 & 20 \\
{Giveme} & 91500 &  30000 & 5 & 5 \\
\bottomrule
\end{tabular}
\label{tab:datatable}
\end{table}
In all experiments, we choose the shift parameter $s_j=-10$ and the truncation parameter $\alpha=20$. 

Note that the target of our scheme is to optimizes the cost of the encryption-related operations in XGBoost. We will study the impacts of three aspects, i.e., secure key length, training data size, and the number of regression trees. The secure key length determines the cost of an encryption operation and size of a cipheretext. The training data size and the number of regression trees will affect the number of encryption operations and ciphertext transmissions. 

\subsection{Performances with Varying Lengths of Secure Key}
\begin{figure}[h]
	\centering
	\subfloat[]{\includegraphics[width=0.8\columnwidth]{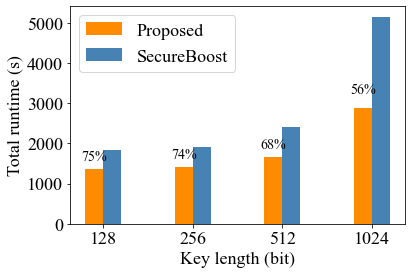}\label{fig:keylen_a}}\hfil
	\subfloat[]{\includegraphics[width=0.8\columnwidth]{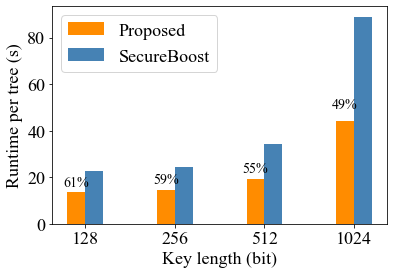}\label{fig:keylen_b}}
	\caption{Comparisons of efficiency with different key lengths. a) The total runtime. b) The average runtime of building per tree.}
	\label{fig:keylen}
\end{figure}
\begin{table}[h]
    \centering
    \caption{Comparisons of accuracy with different key lengths}
    \begin{tabular}{cccccc}
    \toprule
     \multirow{2}{*}{Key length}& \multirow{2}{*}{Algorithm} & \multicolumn{2}{c}{AUC} & \multicolumn{2}{c}{KS} \\
     & & Train & Test &  Train & Test\\
     \midrule
     \multirow{2}{*}{128} &SecureBoost & 0.7900 & 0.7880 & 0.4412 & 0.4408 \\
    &Proposed & 0.7906 & 0.7877 & 0.4434 & 0.4441  \\
    \midrule
     \multirow{2}{*}{256} &SecureBoost & 0.7904 & 0.7875 & 0.4417 & 0.4429  \\
    &Proposed & 0.7899 & 0.7882 & 0.4411 & 0.4435  \\
    \midrule
     \multirow{2}{*}{512} &SecureBoost & 0.7904 & 0.7877 & 0.4404 & 0.4418   \\
    &Proposed & 0.7897 & 0.7866 & 0.4406 & 0.4394 \\
    \midrule
     \multirow{2}{*}{1024} &SecureBoost & 0.7904 & 0.7875 & 0.4417 & 0.4429  \\
    &Proposed & 0.7906 & 0.7877 & 0.4434 & 0.4441  \\
    \bottomrule
    \end{tabular}
    \label{tab:keylen}
    \end{table}
We firstly compare the performance of proposed batch method with SecureBoost on the ``Default'' dataset by varying the key lengths from 128 to 1024. The efficiency of each algorithm is measured by the total running time and the average time of building per tree model. The accuracy of each algorithm is measured by the Area Under Curve (AUC) values and Kolmogorov-Smirnov (KS) values. We set quantization parameter $\alpha_{\sf{max}}=10^6$ and $r=30$. The experimental results in Table \ref{tab:keylen} demonstrate that the accuracy between two algorithms is comparable, while the efficiency shown in Figure \ref{fig:keylen} is considerably improved by our proposed batch method. The number in percentage represents the ratio of ``Proposed'' over ``SecureBoost''. We can observe that the proposed batch method saves more time when the key length increases. The building time of per tree approaches $50\%$ compared with SecureBoost, while the total runtime saves more than $25\%$.

\subsection{Performances with Varying Sample Size}
\begin{figure}[h]
	\centering
	\subfloat[]{\includegraphics[width=0.8\columnwidth]{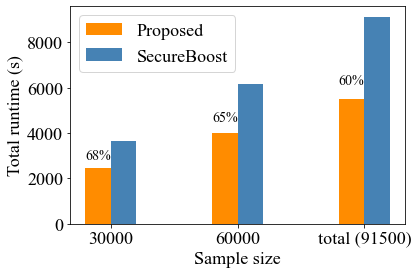}\label{fig:samplesize_a}}\hfil
	\subfloat[]{\includegraphics[width=0.8\columnwidth]{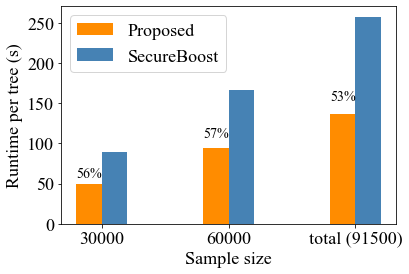}\label{fig:samplesize_b}}
	\caption{Comparisons of efficiency with different sample sizes. a) The total runtime. b) The average runtime of building per tree.}
	\label{fig:samplesize}
\end{figure}
\begin{table}[h]
    \centering
    \caption{Comparisons of accuracy with different sample sizes}
    \begin{tabular}{cccccc}
    \toprule
     \multirow{2}{*}{Sample size}& \multirow{2}{*}{Algorithm} & \multicolumn{2}{c}{AUC} & \multicolumn{2}{c}{KS} \\
     & & Train & Test &  Train & Test\\
     \midrule
     \multirow{2}{*}{30000} &SecureBoost & 0.8596 & 0.8444 & 0.5641 & 0.5455  \\
    &Proposed & 0.8596 & 0.8444 & 0.5625 & 0.5466  \\
    \midrule
     \multirow{2}{*}{60000} &SecureBoost & 0.8498 & 0.8436 & 0.5511 & 0.5401   \\
    &Proposed & 0.8499 & 0.8436 & 0.5526 & 0.5401  \\
    \midrule
     \multirow{2}{*}{Total (91500)} &SecureBoost & 0.8533 & 0.8436 & 0.5587 & 0.5399  \\
    &Proposed & 0.8533 & 0.8438 & 0.5577 & 0.5407 \\
    \bottomrule
    \end{tabular}
    \label{tab:samplesize}
    \end{table}
We then conduct numerical experiments on the ``Giveme'' dataset to show the performances under different sample sizes. We sample $30000$ and $60000$ data instances from the training set uniformly at random, and run each algorithm on the dataset of different size. We set quantization parameter $\alpha_{\sf{max}}=10^7$ and $r=40$ since the possible number of aggregation gradients are much larger. The accuracy results are presented in Table \ref{tab:samplesize} and the efficiency results are illustrated in Figure \ref{fig:samplesize}. The numerical results demonstrate that the proposed batch method enjoys more efficiency improvement when the problem size increases, and the total runtime has been saved more than $30\%$. By choosing proper parameters, the performance of the proposed batch method is still be comparable to SecureBoost. 

\subsection{Performances with Varying Number of Trees}
\begin{figure}[h]
	\centering
	\subfloat[]{\includegraphics[width=0.8\columnwidth]{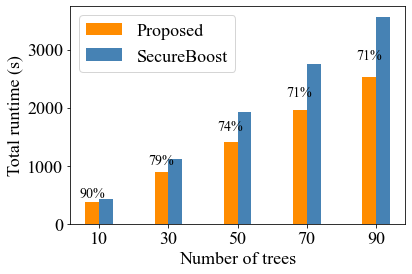}\label{fig:round_a}}\hfil
	\subfloat[]{\includegraphics[width=0.8\columnwidth]{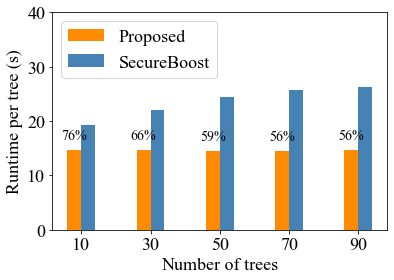}\label{fig:round_b}}
	\caption{Comparisons of efficiency with different number of boosting trees. a) The total runtime. b) The average runtime of building per tree.}
	\label{fig:round}
\end{figure}
\begin{table}[h]
    \centering
    \caption{Comparisons of accuracy with different number of boosting trees}
    \begin{tabular}{cccccc}
    \toprule
     \multirow{2}{*}{Boosting round}& \multirow{2}{*}{Algorithm} & \multicolumn{2}{c}{AUC} & \multicolumn{2}{c}{KS} \\
     & & Train & Test &  Train & Test\\
     \midrule
     \multirow{2}{*}{10} &SecureBoost & 0.7671 & 0.7722 & 0.4170 & 0.4260   \\
    &Proposed & 0.7664 & 0.7707 & 0.4164 & 0.4246  \\
    \midrule
     \multirow{2}{*}{30} &SecureBoost & 0.7791 & 0.7836 & 0.4295 & 0.4436 \\
    &Proposed & 0.7804 & 0.7851 & 0.4333 & 0.4441   \\
    \midrule
     \multirow{2}{*}{50} &SecureBoost & 0.7904 & 0.7875 & 0.4417 & 0.4429  \\
    &Proposed & 0.7899 & 0.7882 & 0.4411 & 0.4435 \\
    \midrule
     \multirow{2}{*}{70} &SecureBoost & 0.7967 & 0.7894 & 0.4501 & 0.4488  \\
    &Proposed & 0.7967 & 0.7906 & 0.4501 & 0.4574 \\
    \midrule
     \multirow{2}{*}{90} &SecureBoost & 0.8021 & 0.7904 & 0.4557 & 0.4525 \\
    &Proposed & 0.8025 & 0.7909 & 0.4561 & 0.4514  \\
    \bottomrule
    \end{tabular}
    \label{tab:round}
    \end{table}
In this subsection, we evaluate the performances of each algorithm by building different number of trees, i.e., the boosting round, on the ``Default'' dataset. We set the number of boosting rounds as $10,30,50,70,90$, the quantization parameter $\alpha_{\sf{max}}$ as $10^6$ and $r=30$. The accuracy results are provided in Table \ref{tab:round} and the efficiency results are illustrated in Figure \ref{fig:round}. The numerical results demonstrate that the efficiency advantage for the proposed batch method is bigger when the number of boosting rounds increases while remaining comparable accuracy.

Through above numerical experiments, we find that the total runtime with the proposed method is from $75\%$ to $55\%$ of the time with SecureBoost, while the time of building boosting trees is only from $60\%$ to $50\%$. The precision loss during encoding of the proposed method does not degrade the performance of the trained vertical federated XGBoost model (difference no greater than). 

\section{Conclusion}
In this paper, we studied vertical federated learning, in particular, XGBoost modeling. To address the limitations of inefficiency resulted by tremendous homomorphic encryption operations and transmissions, we proposed a batch method to encode the first-order and second-order derivatives into a single number, which cuts the cost of encryption and ciphertext transmission almost in half. The proposed method avoided the overflow caused by adding negative numbers when directly applying existing batching method. We then provided theoretical analysis on the selection of hyperparameters and their effects on possible overflow and encoding precision. Numerical experiments demonstrated the much higher efficiency and comparable accuracy (measured by AUC and KS) of our method.

\bibliographystyle{IEEEtran.bst}
\bibliography{batch}

\end{document}